\documentclass[aip,jcp,amsmath,amssymb,superscriptaddress]{revtex4-1}

\usepackage{graphicx}
\usepackage{bm}

\newcommand{\be}{\begin{equation}} 
\newcommand{\ee}{\end{equation}} 
\newcommand{\bea}{\begin{eqnarray}}
\newcommand{\eea}{\end{eqnarray}}
\newcommand{\nn}{\nonumber}

\newcommand{\onehalf}{\frac{1}{2}}
\newcommand{\kBTotwo}{\frac{k_BT}{2}}

\newcommand{\RR}{\mathbf{r}}
\newcommand{\rr}{\mathbf{r}}

\newcommand{\dr}{d\mathbf{r}}

\newcommand{\kk}{\mathbf{k}}

\newcommand{\rhon}{{n\left({\mathbf r}\right)}}

\newcommand{\rhorom}{{\rho\left({\mathbf r},\Omega\right)}}
\newcommand{\alpharom}{{\alpha\left({\mathbf r},\Omega\right)}}
\newcommand{\Om}{\mathbf{\Omega}}

\newcommand{\nr}{n(\mathbf{r})}

\newcommand{\F}{{\cal F}}

\newcommand{\Pol}{\mathbf{P}({\mathbf r})}

\newcommand{\MU}{\boldsymbol{\mu}}
\newcommand{\rhonbar}{\bar{n}({\mathbf r})}
\newcommand{\nbar}{\bar{n}}

\newcommand{\PP}{\mathbf{P}}
\newcommand{\EE}{\mathbf{E}}

\newcommand{\etal}{{\em et  al. }}

\newcommand{\angstrom}{\r{a}ngstr\"{o}m}

\begin{document}

\title{Molecular Density Functional Theory of Water describing Hydrophobicity at Short and Long Length Scales}

%~ \pacs{}
%~ \keywords{}

\author{Guillaume Jeanmairet}
\affiliation{P\^ole de Physico-Chimie Th\'eorique, \'Ecole Normale Sup\'erieure, UMR 8640 CNRS-ENS-UPMC, 24, rue Lhomond, 75005, Paris, France}

\author{Maximilien Levesque}
\email{maximilien.levesque@ens.fr}
\affiliation{UPMC Univ Paris 06, UMR 7195, PECSA, F-75005, Paris, France}
\affiliation{CNRS, UMR 7195, PECSA, F-75005, Paris, France}

\author{Daniel Borgis}
\email{daniel.borgis@ens.fr}
\affiliation{P\^ole de Physico-Chimie Th\'eorique, \'Ecole Normale Sup\'erieure, UMR 8640 CNRS-ENS-UPMC, 24, rue Lhomond, 75005, Paris, France}

\begin{abstract}
We present an extension of our recently introduced molecular density functional theory of water [G. Jeanmairet \etal, J. Phys. Chem. Lett. {\bf 4}, 619, 2013] to the solvation of hydrophobic solutes of various sizes, going from \angstrom s to nanometers. The theory is based on the quadratic expansion of the excess free energy in terms of two classical density fields, the particle density and the multipolar polarization density. Its implementation requires as input a molecular model of water and three measurable bulk properties, namely the structure factor and the $k$-dependent longitudinal and transverse dielectric susceptibilities. The fine three-dimensional water structure around small hydrophobic molecules is found to be well reproduced. In contrast the computed solvation free-energies appear overestimated and do not exhibit the correct qualitative behavior when the hydrophobic solute is grown in size. These shortcomings are corrected, in the spirit of the Lum-Chandler-Weeks theory, by complementing the functional with a truncated hard-sphere functional acting beyond quadratic order in density, and making the resulting functional compatible with the Van-der-Waals theory of liquid-vapor coexistence at long range. Compared to available molecular simulations, the approach yields reasonable solvation structure and free energy of hard or soft spheres of increasing size, with a correct qualitative transition from a volume-driven to a surface-driven regime at the nanometer scale.
\end{abstract}

\maketitle

\section{Introduction}

The numerical  methods that have emerged in the second part of the last century from  liquid-state theories\cite{hansen, gray-gubbins-vol1}, including integral equation theory in the interaction-site\cite{Chandler-RISM,hirata-rossky81,hirata-pettitt-rossky82,reddy03,pettitt07,pettitt08}
or molecular\cite{blum72a,blum72b,patey77,fries-patey85,richardi98,richardi99} picture, classical density functional theory (DFT)\cite{evans79,evans92,Wu07}, or classical fields theory\cite{chandler93,lum99,coalson96}, have become methods of choice for 
many physical chemistry or chemical engineering applications\cite{gray-gubbins-vol2,neimark06,neimark11,wu06}. They can yield reliable predictions for both the microscopic structure and the thermodynamic properties of molecular fluids in bulk, interfacial, or confined conditions at a much more modest computational cost than  molecular dynamics or Monte-Carlo simulations. 
A current challenge concerns their implementation in three dimensions in order to describe molecular liquids, solutions, and mixtures in  complex environments such as atomistically-resolved solid interfaces or biomolecular media. There have been a number of recent efforts in that direction using 3D-RISM\cite{Beglov-Roux97,kovalenko-hirata98,red-book,yoshida09,kloss08-jcp,kloss08-jpcb} or site DFT\cite{liu13}, lattice field\cite{azuara06,azuara08} or Gaussian field\cite{lum99,tenwolde01,tenwolde02,huang02,varilly11,chandler-varilly11} theories.
Recently, a molecular density functional theory (MDFT) approach to solvation has been
introduced.~\cite{ramirez02,ramirez05-CP,ramirez05,gendre09,zhao11,borgis12,levesque12_1,levesque12_2,jeanmairet13} It relies on the definition of a free-energy functional depending on the full six-dimensional position and orientation solvent density. In the so-called homogeneous reference fluid (HRF) approximation, the (unknown) excess free energy can be inferred from the angular-dependent direct correlation function of the bulk solvent,
that can be predetermined from molecular simulations of the pure solvent. Compared to reference molecular dynamics calculations, such approximation was shown to be accurate for polar, non-hydrogen bonded fluids \cite{ramirez02,gendre09,zhao11,borgis12,levesque12_2}, and for water \cite{zhao11,jeanmairet13,levesque12_1}, if appropriate many-body corrections are introduced to either strengthen the tetrahedral order around H-bonding sites\cite{zhao11,jeanmairet13}, or improve the solvation free-energies of small hydrophobic solutes\cite{levesque12_1}.

In a recent work\cite{jeanmairet13}, we have introduced a simplified version of MDFT that can be derived rigorously for simple point charge representations of water such as SPC or TIP4P, involving a  single Lennard-Jones interaction site and distributed partial charges. In that case we showed that the functional can be expressed in terms of the particle density $n(\rr)$ and site-distributed polarisation density $\PP(\rr)$, and it requires  as input three simple bulk physical properties of water, namely the density structure factor $S(k)$, and the $k$-dependent longitudinal and transverse dielectric susceptibilities, $\chi_L(k)$ and $\chi_T(k)$, which are related to the longitudinal and transverse dielectric constants $\epsilon_L(k)$ and $\epsilon_T(k)$\cite{raineri92,raineri93,bopp98}. Those quantities can be inferred from experiments, or from molecular dynamics simulations of the selected point-charge model in bulk conditions\cite{bopp96,bopp98}. Elaborating on Refs.~\onlinecite{levesque12_1,jeanmairet13}, we  show in this paper how the  functional can be corrected on the basis of physical arguments  in order to describe the solvation of hydrophobic solutes of various sizes, going from microscopic to mesoscopic length scales. The approach is largely inspired by the theory of hydrophobic hydration developed by Lum, Chandler and Weeks (LCW)\cite{lum99}, and later by Chandler and collaborators
\cite{tenwolde01,tenwolde02,huang02,varilly11,chandler-varilly11}, pointing the fact that the solvation of hydrophobic solutes has a different physical nature at those different  scales.  Hughes {\em et al.} have recently proposed a classical DFT approach relying on the hard-sphere fundamental measure theory (FMT) plus attractive interactions based on the
statistical associating fluid theory  (SAFT)  to address the same problem\cite{roundy13}. In addition,  our theory is also able to account for the self-consistent coupling of the density field to the  polarization field. It will be used to compute the hydration properties of various molecular solutes that can be classified as hydrophobic, and can be described by  Lennard-Jones site parameters and  either a vanishing or small partial charge distribution on the sites.

This paper is organized as follows. Sec.~2 presents our extension of MDFT that describes  hydrophobic effects at short and long length scales. Sec.~3  gives details on the practical numerical implementation. Sec.~4 shows a few applications to the hydration properties of solutes of varying size. Sec.~5 concludes.

\section{Theory}

In Ref.~\onlinecite{jeanmairet13}, we showed that the Helmholtz free-energy of a liquid water sample  in the presence of an embedded solute  creating at each point in space a Lennard-Jones potential
$\Phi_{LJ}(\rr)$ and an electric field  $\mathbf{E}_c(\rr)$, can be expressed as a functional of two fields, the particle number density $\nr$ and the polarization density, $\Pol$, instead of the more general functional of  the position and orientation density, $\rhorom$, which applies for an arbitrary solvent\cite{gendre09,zhao11,borgis12}. Those fields can be defined from $\rhorom$ by
\bea
\nr & = &\int d\Omega  \, \rhorom \\
 \Pol   & = & \int d\rr' d\Omega \, \MU(\rr-\rr', \Omega) \rho(\mathbf{r}',\Omega),
 \eea
where $d\Omega$ indicates an angular integration over the three Euler angles and $\MU(\rr,\Omega)$ is the polarization density of a single water molecule placed at the origin. According to the definition of Raineri \etal \cite{raineri93},  this quantity is better apprehended
in $k$-space as 
\bea
\MU(\kk,\Omega) &= & -i \, \sum_m q_m \frac{\mathbf{s}_m(\Omega) }{\kk \cdot \mathbf{s}_m(\Omega) } \left[ e^{i \, \kk \cdot \mathbf{s}_m(\Omega) } -1 \right] \\
\label{eq:mukom}
&= & \MU(\Omega) + \frac{i}{2} \sum_m q_m \left[ \kk \cdot \mathbf{s}_m(\Omega) \right]  \mathbf{s}_m(\Omega) + ... 
\eea
where $q_m$ and $s_m(\Omega)$ stand for the partial charges and coordinates of site $m$ for a fixed molecule orientation.
 In the second equality, the first term $\MU(\Omega) = \sum_m q_m \, \mathbf{s}_m(\Omega) =  \mu \, \Om$  is the orientation-dependent molecular dipole (with $\mu$ being the  dipole magnitude; for water,
 $\Om$  is the unit vector along the $C_{2}$ axis). The following terms account for the multipolar expansion of the charge density, starting from the quadrupole. 

We thus start from the functional  $\F[n(\rr),\Pol]$ introduced in Ref.~\onlinecite{jeanmairet13} for liquid water in an external field.  Expressing the ideal term as a functional of the whole position and orientation density, $\rho(\rr,\Omega)$, instead of the unknown expression in terms of just $n(\rr)$ and  $\PP(\rr)$ (an approximation reminiscent of the Kohn-Sham approximation for the kinetic energy in electronic DFT),  one is lead to the following total functional 
\bea
\label{eq:Ftot}
\F[n(\rr),\PP(\rr)] &=& k_BT \int d\rr d\Omega \left[ \rhorom \ln \left(\frac{8\pi^2 \rhorom}{n_0} \right) - \rhorom+\frac{n_0}{8\pi^2} \right] \nn \\
& + &   \int \dr \, \Phi_{LJ}(\rr) n(\rr) - \int \dr \, \EE_c(\rr) \cdot \PP(\rr) + \F_{exc}[n(\rr),\PP(\rr)],
\eea
where $n_0$ is the bulk solvent density. Neglecting the direct correlation term between $\nr$ and $\Pol$ in the excess functional, the above expression can be easily decomposed into a density (translational) and polarization (orientational) contribution, {\em i.e}
\be
  \F[n(\rr),\PP(\rr)] = \F_n[\rhon] + \F_P[\rhon,\alpharom]
  \label{eq:Fn+FP}
\ee
with
\bea
\F_n[\rhon] &= &k_BT \,\int d\RR \,  \left[ \rhon \ln(\frac{\rhon}{n_0})-\rhon+n_0 \right]  \nn \\
& & + \int d\rr\, \rhon \,  \Phi_{LJ}(\rr) + \F_n^{exc}[n(\rr)]   \label{eq:Fn}
\eea
and
\bea
\F_P[\rhon,\alpharom] &=& k_BT \int d\rr  \, \rhon \int d\Om \left[ \alpharom \ln \left(\frac{\alpharom}{\alpha_0} \right)  - \alpharom + \alpha_0 \right] \nn \\
& &  - \int \dr \, \EE_c(\rr) \cdot \PP(\rr) + \F_P^{exc}[\Pol], 
\label{eq:FP}
\eea
where we have used  the conditional orientational probability $\alpharom = \rhorom/\rhon$. In this last equation, the excess free energy is expressed at quadratic order around the homogeneous fluid state
\bea
\F_P^{exc}[\Pol] &= & -  \frac{3k_BT}{2\mu^2 n_0}  \int d\rr \, \Pol^2\nn \\
&+&      \frac{1}{8\pi \epsilon_0} \int d\rr_1 d\rr_2 \, \chi_T^{-1}(r_{12};n_0) \, \PP_T(\rr_1) \cdot \PP_T(\rr_2)  \nn \\
&+&    \frac{1}{8\pi \epsilon_0} \int d\rr_1 d\rr_2 \, \chi_L^{-1}(r_{12};n_0) \, \PP_L(\rr_1) \cdot \PP_L(\rr_2).  
\label{eq:FP_exc}
\eea
This expression involves the longitudinal and transverse components of the polarization vector $\Pol$, defined in $k$-space by
\be
\PP_L(\kk) = \left(\PP(\kk) \cdot \hat{\kk}\right)  \hat{\kk}, \,  \, \PP_T(\kk) = \PP(\kk) - \PP_L(\kk),
\label{eq:P_L_T}
\ee
as well as their associated longitudinal and transverse linear response susceptibilities, which are density-dependent bulk fluid properties. Those quantities can be extracted from experiments, or from molecular simulation using a given water model; they are reported in 
Ref.~\onlinecite{jeanmairet13} for SPC/E water.
%we have computed them  and we display them in Fig.~\ref{fig:chi_L_and_T} for SPC/E water. 

Note here that for a purely hydrophobic solute carrying no charges, $\EE_c = 0$. 
$\F_P[\rhon,\alpharom]$ is then a purely quadratic functional of $\alpharom$ which yields $\alpharom \equiv 0$ by minimization. 
We are left in this case  with the particle-density functional $\F_n[\rhon]$ of eq.~\ref{eq:Fn}.  In the general case where the external electric field is non-zero, the coupling between density and polarisation density occurs through the ideal (entropic) term of eq.~\ref{eq:FP}.

As for the density-dependent excess functional in eq.~\ref{eq:Fn},  we write
\bea
\F_n^{exc}[n(\rr)] &=&  -\frac{k_BT}{2n_0}  \int d\rr \Delta n(\rr)^2 + \frac{k_BT}{2}  \int d\rr_1 d\rr_2 \, \left[S^{-1}(r_{12};n_0) \, \Delta n(\rr_1) \,  \Delta n(\rr_2)\right]  + \F_B[\rhon] \nn \\
&=& - \frac{k_BT}{2} \int d\rr_1 d\rr_2 \, \left[c_S(r_{12};n_0) \, \Delta n(\rr_1) \,  \Delta n(\rr_2)\right]  + \F_B[\rhon],
\label{eq:Fn_exc} 
\eea
where $r_{12}=|\bm{r_1-r_2}|$ and $\Delta n (\bm{r})=n(\bm{r})-n_0$.
The first terms that are quadratic in $\Delta n(\rr)$ correspond  to a second order expansion around the homogeneous fluid state (the so-called homogeneous reference (HRF) approximation \cite{ramirez02}) and they involve either the  inverse of the density susceptibility or the isotropic component of the water direct correlation function (In $k$-space: $n_0c_S(k;n_0) = 1 - S^{-1}(k;n_0)$). The latter function was extracted from the radial distribution function $g(r)$  using the Baxter direct-space method, assuming $c(r)=0$ beyond $r = 9$~\AA \cite{ramirez05-CP}. 
In eq.~\ref{eq:Fn_exc}, all   terms of  orders in $\Delta n$ that are higher than 2 are gathered in the  so-called bridge functional. As in Ref.~\onlinecite{levesque12_1} and  in many others works\cite{rosenfeld93,zhao-wu11,zhao-wu11-correction,liu13,oettel_2005}, we here approximate the bridge term by a hard-sphere bridge (HSB) for a reference harde-sphere fluid of density $n_0$ and radius $R_0$
\bea
\F_B[\rhon] &= &\F_{exc}^{HS}[\rhon] -   \F_{exc}^{HS}[n_0] - \mu_{exc}^{HS} \int d\rr \Delta\rhon \nn \\
& +  & \frac{k_BT}{2} \int d\rr_1d\rr_2 \Delta n(\rr_1) \, \Delta n(\rr_2) \, c_S^{HS}(r_{12};n_0)  
\label{eq:Fn_bridge}
\eea
involving the hard-sphere direct correlation function  and chemical potential  at the density $n_0$. 
This bridge term starts indeed with an  order $\Delta n^3$.

\begin{figure}
    \includegraphics[width=8.5cm]{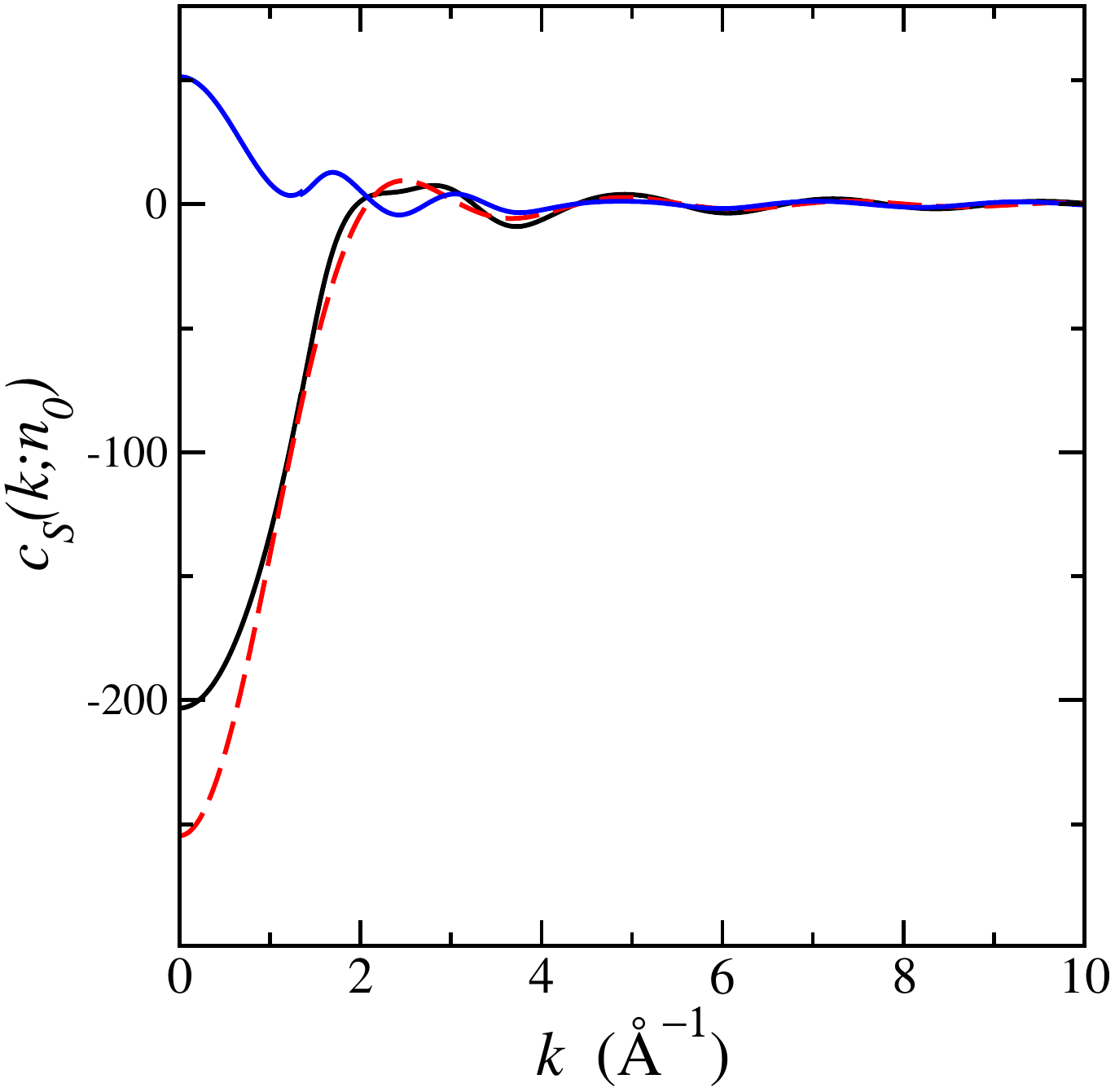}
    \caption{Comparison of direct correlation functions of water and hard sphere fluid:\\
    Direct correlation function $c_S(k;n_0)$ for water (black) and a hard-sphere fluid with $R_0= 1.27$~\AA~ (red) at identical density  $n_0=0.333$~molecules/\AA$^3$. The upper blue curve is their difference, i.e. the "attractive" direct correlation function $c_S^a(r;n_0)$ defined in the text. This curve does verify the conditions $a=c_S^a(k=0)  >0$ and $m = -d^2c_S^a/dk^2(0) > 0$.
        \label{fig:ck_spc}
        }
\end{figure}

Furthermore, as Lum, Chandler, and Weeks\cite{lum99}, we further define a coarse-grained density $\rhonbar$ by
\be 
\rhonbar = \int d\rr_2 \, G(r_{12}) \, n(\rr_2)
\label{eq:rhobar}
\ee
or in $k$-space
\be
\nbar(\kk) = G(k) n(\kk)
\label{eq:14}
\ee
where $G(r)$ is a coarse-graining function. At the moment, let us suppose for the purpose of the derivation that $G(k) = \Theta(k_c-k)$, i.e equal 1 for $k < k_c$ and 0 otherwise ($k_c$ being typically of the order of a few \AA$^{-1}$). In that case 
\be
\nbar(\kk) (n(\kk) - \nbar(\kk) ) = 0 \, \, \textrm{for all}\,  \kk.
\label{eq:n-nbar}
\ee
The  terms in eqs~\ref{eq:Fn_exc}-\ref{eq:Fn_bridge} that are quadratic in $\Delta n$ can be gathered to define an attractive free-energy term, reminescent of the Van-der-Waals theory, and defined by
\be
\F_{n,a}^{exc}[\rhon]  =  -\kBTotwo   \int d\rr_1d\rr_2  \, c_S^{a}(r_{12}) \,  \Delta n(\rr_1) \, \Delta n(\rr_2) 
\ee
with $c_S^{a}(r_{12}) =  c_S(r_{12};n_0) - c_S^{HS}(r_{12};n_0)$. According to the assumed property in eq.~\ref{eq:n-nbar}, this term can be further decomposed into a short-range and long-range term
\bea
\F_{n,a}^{exc}[\rhon] & = &  -\frac{k_BT}{2} \int d\rr_1 d\rr_2 \, c_S^{a}(r_{12}) \, (n(\rr_1) - \nbar(\rr_1)) \,  (n(\rr_2) - \nbar(\rr_2)) \nn \\
& &  -\kBTotwo \int d\rr_1d\rr_2 \, c_S^{a}(r_{12}) \, \Delta \nbar(\rr_1) \,  \Delta \nbar(\rr_2).
\eea
 Using the prescription of Evans\cite{evans79,evans92} that $c_S^a(r)$ is short ranged, whereas $\nbar(\rr)$ is a coarse-grained, slowly varying function, the second, long-ranged integral in the above equation can be expressed in $k$-space and approximated as 
\bea
\int d\kk \, c_S^a(k) \, \Delta \nbar(\kk) \, \Delta \nbar(-\kk) &= & \int d\kk  \left(  c_S^a(k=0) + k^2 \frac{d^2 c_S^a}{dk^2}(k=0) + ... \right) \Delta \nbar(\kk) \, \Delta \nbar(-\kk) \nn \\
& = &   a \int d\kk \Delta \nbar(\kk) \, \Delta \nbar(-\kk) - m \int d\kk k^2 \Delta \nbar(\kk) \, \Delta \nbar(-\kk) + ...  \nn \\
&= &  a \int d\rr \Delta\nbar(\RR)^2 -  m \int d\rr \left( \mathbf{\nabla}  \nbar(\RR) \right)^2  + ...
\label{eq:F_exc_expansion}
\eea
where $a$ and $m$ are positive parameters (as can be checked in Fig.~\ref{fig:ck_spc}) that can be related to the attractive energy in Van-der-Waals theory and the liquid-gas surface tension, $\gamma$. They will be considered hereafter as two phenomenological, adjustable parameters. Summarizing the above developments, the density-dependent excess free-energy in eq.~\ref{eq:Fn} can be written as
\bea \label{eq:Fn_tot}
\F_n^{exc}[n(\rr)] &= & - \frac{k_BT}{2} \int d\rr_1 d\rr_2 \, \left( c_S(r_{12};n_0)- c_S^{HS}(r_{12};n_0) \right) \, (n(\rr_1) - \nbar(\rr_1)) \,  (n(\rr_2) - \nbar(\rr_2)) \nn \\
&+& \kBTotwo \int d\rr \left( -a  \Delta \rhonbar^2 + m \left( \mathbf{\nabla}  \nbar(\RR) \right)^2 \right) \nn \\
& + &\F_{exc}^{HS}[\rhon] -  \F_{exc}^{HS}[n_0] - \mu_{exc}^{HS} \int d\rr \Delta \rhon.
\label{eq:Fn_exc_2}
\eea
For the hard-sphere functional, we invoke here the scalar fundamental measure theory functional of Kierlik and Rosinberg (KR-FMT)\cite{kierlik-rosinberg90,kierlik-rosinberg91}, which was shown to be equivalent to the vectorial form of Rosenfeld\cite{rosenfeld89,phan93}, and which, as we showed in Ref.~\onlinecite{levesque12_1},  is easily and efficiently implementable in three dimensions with fast Fourier transforms. For the slowly varying density $\rhon = \rhonbar$, the short-range term vanishes in eq.~\ref{eq:Fn_exc_2} and the hard-sphere excess functional can be assumed to adopt its local, Percus-Yevick or Carnahan-Starling form, according to the KR functional chosen\cite{levesque12_1}:
\be
\F_{exc}^{HS}[\rhonbar]  = \int d\rr \, \rhonbar f_{exc}^{HS}(\rhonbar),
\ee
with in the case of the Carnahan-Starling choice
\be
f_{exc}^{HS}(n)= \frac{\eta(4 - 3 \eta)}{(1-\eta)^2}, \,
\ee
with $\eta = 4 \pi R_0^3 n/3$ the packing fraction and $R_0$ the hard-sphere diameter. Compared to PY, the CS choice provides a more precise expression of the homogeneous free energy, but at the expense of some thermodynamics inconsistencies\cite{kierlik-rosinberg90}. To our experience, both choices yield close results for liquid conditions\cite{levesque12_1}.  The intrinsic part of $\F[\rhonbar]$ (ideal plus excess parts) can then be written as 
\be 
\F_{n}^{int}[\rhonbar] = k_BT \int d\rr \, \left[ F_{VdW}(\rhonbar) + \onehalf m \left( \mathbf{\nabla}  \nbar(\RR) \right)^2 \right]
\label{eq:F_n_int}
\ee 
with
\bea
F_{VdW}(\rhonbar) & = &   \rhonbar \ln\left(\frac{\rhonbar}{n_0}\right)- \Delta \rhonbar
+ \rhonbar f_{exc}^{HS}(\rhonbar) -  n_0 f_{exc}^{HS}(n_0) \nn \\
&-& \left( f_{exc}^{HS}(n_0)  + n_0 \frac{df_{exc}^{HS}}{dn}(n_0) \right) \Delta \rhonbar - \onehalf a \Delta \rhonbar^2.
\label{eq:F_VdW}
\eea
Since liquid water at ambient condition lies very close to the liquid-gas coexistence, a first clue to select the Van-der-Waals parameter $a$ is to assume near equality of the gas and liquid free-energies, i.e. to impose that $F_{VdW}(n)$ has two equivalent minima, one at $n_0$ and the other at a much lower density pertinent to the gas phase. Two such curves are represented in Fig.~\ref{fig:fn} for $n_0=0.0333$ molecules/\AA$^3$, the water density at ambient conditions, and two choices of the hard-sphere radius: $R_0=1.25$~\AA, the value corresponding to the hard core in the radial distribution function, and $R_0 = 1.4$~\AA, that can be inferred from  the nearest neighbor distance in water. Both values were used in the literature\cite{oleksy10,zhao-wu11,levesque12_1} to account for the hard core of water. For those two values, the condition of  iso-energetic minima is met for  $n_0a = 6.7$ or $12.5$, respectively. For the smallest radius $R_0 = 1.25$~\AA, the gas-phase density appears slightly too high:  The minimal parametrization adopted here does not allow to impose the value of the gas phase density nor that of the barrier height. In all cases however, the key physical feature is the near co-existence of a low density and high density phase. As for the choice of the parameter  $m$ in eq.~\ref{eq:F_n_int}, it  appears less critical and and we took $n_0m=1.0$~\AA$^2$.
The phenomenological parameters $a$ and $m$ being prescribed, the final functional to be minimized for a given external Lennard-Jones and Coulomb field is given by eqs.~\ref{eq:Fn}-\ref{eq:FP_exc} and~\ref{eq:Fn_exc_2}. We note that the theory developed here is fully self-consistent: for the density part, $\F_n[n(\rr)]$ is written  as a functional of $\rhon$ and $\rhonbar$ but the minimization is eventually performed with respect to  $\rhon$ since $\rhonbar$ is  linked to $\rhon$ through eq.~{\ref{eq:rhobar}}. This is done as follows.

\section{Functional minimization}

In the general case, the total functional  is minimized with respect to $\rhorom$, that is discretized  on a  position and orientation grid.  The minimization is performed in fact with respect to a ''fictitious wave function'', $\Psi(\rr,\Omega)=\rho(\rr,\Omega)^{1/2}$,  in order to prevent the density from becoming negative in the logarithm. The position are represented on a 3D-grid with typically 3-4  points per \angstrom \ whereas the orientations are discretized using a Lebedev grid (yielding the most efficient angular quadrature for the unit sphere\cite{lebedev99}) for $\theta, \phi$, the orientation of the molecule $C_{2}$ axis, and a regular grid for the remaining angle $\psi$ (from $0$ to $\pi$). We use typically $N_\Omega = 12 \times 3$ angles. The calculation begins with the tabulation of the external Lennard-Jones potential $\phi_{LJ}(\rr)$, and external electric field $\EE_c(\rr)$. The latter is computed by extrapolating the solute charges on the grid and solving the Poisson equation. Concerning the polarization part, the molecular dipole function $\MU(\kk,\Omega)$ is also precomputed in $k$-space.   At each minimization cycle, $\rhorom$ is Fourier transformed to get 
$\rho(\kk,\Omega)$,  which is integrated over angles to get $n(\kk)=\int d\Omega \, \rho(\kk,\Omega)$ and $\PP(\kk) = \int d\Omega \, \MU(\kk,\Omega) \rho(\kk,\Omega)$,  as well as  its longitudinal and transverse components defined by eq.~\ref{eq:P_L_T}.
The non-local excess terms in eq.~\ref{eq:FP_exc} are computed in $k$-space and so are the gradients with respect to $\rho(\kk,\Omega)$; those are then back-transformed to r-space. 

The density-dependent terms in eqs~\ref{eq:Fn}-\ref{eq:Fn_tot}  that depend only on $n(\rr)$ and 
$\rhonbar$ can be differentiated with respect to $\rhorom$ according to 
\begin{equation}
    \frac{\delta \F_n}{\delta \rhorom} =  \frac{\delta \F_n}{\delta n(\rr)} + \int d\rr' \, \frac{\delta \F_n}{\delta \rhonbar} \, G(|\rr - \rr'|).
\end{equation}
The convolution with the coarse-graining function $G(r)$  is performed in $k$-space and then inverse Fourier transformed. Although the formulas above where derived rigorously for a particular form of $G(r)$, they apply in effect to any other sensible choice of coarse-graining; Following LCW\cite{lum99}, we have taken a Gaussian function with a standard deviation of $4$~\AA.  Finally, for the hard sphere FMT functional of eq.~\ref{eq:Fn_tot}, we used the three-dimensional implementation of the scalar Kierlik-Rosinberg version described in Ref.~\onlinecite{levesque12_1}, with the Carnahan-Starling equation of states. It requires the convolution of the microscopic density $n(\rr)$ by five scalar geometric  weight functions and a total of 10 FFT's per minimization cycle to get the corresponding free energies and gradients. 

For minimization, we used the L-BFGS quasi-Newton optimization routine\cite{BFGS} which requires as input at each cycle free energy and gradients.  The calculation are performed on  a standard  workstation or laptop. The convergence is rather fast, requiring at most 25-30 iterations, so that despite the important number of FFT's handled at each step, one complete minimization takes at most a few minutes on a single processor.  For a purely hydrophobic solute (no charges), the functional becomes independent of angles,  and the minimization requires tens of seconds rather than minutes.

\section{Hydration of hydrophobic solutes of various length scales}

\begin{figure}
    \includegraphics[width=8.5cm]{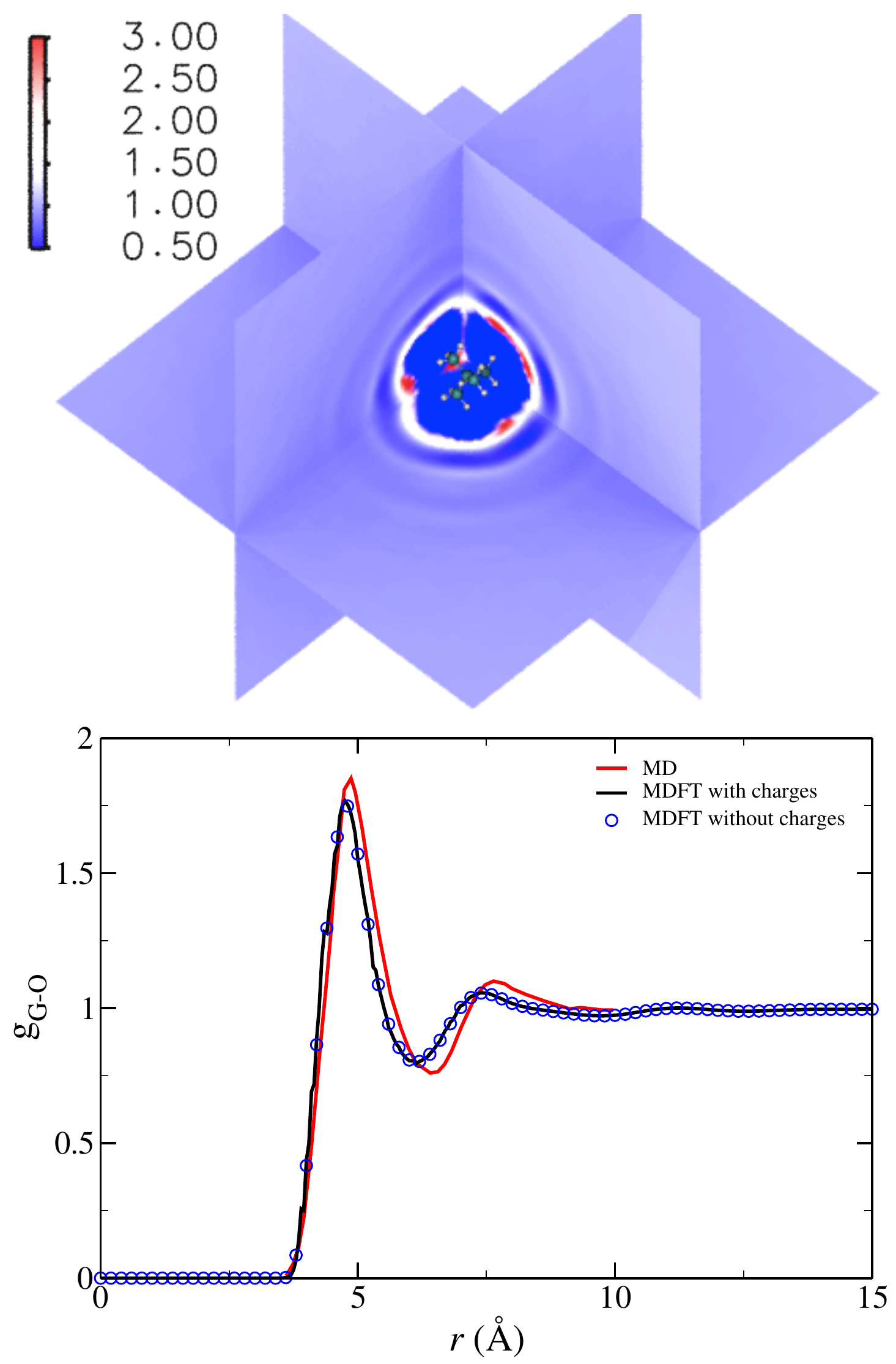}
    \caption{
        \label{fig:gr_neopentane}
        Water stucture around a neopentane molecule:\\At the top, the neopentane molecule and the associated three-dimensional water density obtained by functional minimization. At the bottom, the radial distribution function between the neopentane center of mass and the oxygen of water as calculated by MDFT with hard-sphere bridge and Van-der-Waals correction with atomic partial charges (black line) and without (blue circle). The reference radial distribution function as calculated by MD is shown in red.
        }
\end{figure}

\begin{figure}
   \includegraphics[width=8.5cm]{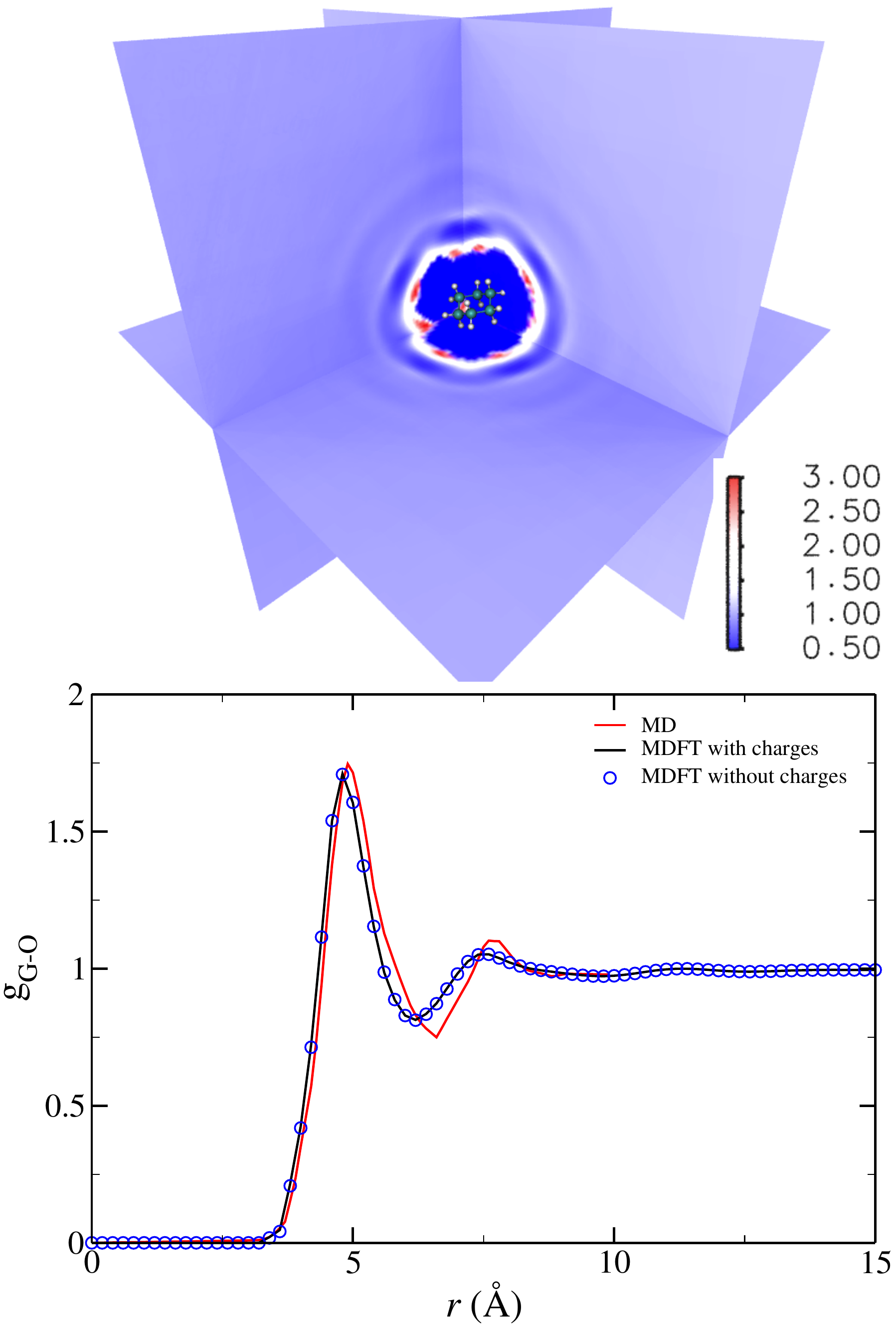}
    \caption{
        \label{fig:gr_cyclohexane}
                Water structure around a cyclohexane molecule:\\At the top, the cyclohexane molecule and the associated three-dimensional water density obtained by functional minimization. At the bottom, the radial distribution function between the cyclohexane center of mass and the oxygen of water as calculated by MDFT with hard-sphere bridge and Van-der-Waals correction with atomic partial charges (black line) and without (blue circle). The reference radial distribution function as calculated by MD is shown in red.
  }
\end{figure}

As an illustration of the formalism above, we begin with the hydration in SPC/E water of two typical small hydrophobic molecules, cyclohexane and neopentane that, in force fields like Optimized Potentials for Liquid Simulations (OPLS), are typically described by a charge distribution in addition to the atomic Lennard-Jones parameters. Those parameters were taken from Refs.~\onlinecite{raschke_2004} and \onlinecite{Huang_2003} respectively. Figures \ref{fig:gr_neopentane} and \ref{fig:gr_cyclohexane} illustrate that the whole MDFT formalism including both the density and polarization fields as variables is able to provide the three-dimensional solvent structure with fine resolution in less than a minute on a standard single processor. We used here a cubic supercell of 40*40*40~\AA$^3$ and $120^3$ grid points. It can be seen also that MDFT minimization yields  microscopic solvent structures, reduced here to the molecule center of mass/water radial distribution function, that compare reasonably well with those obtained from MD simulations with the same molecular models, reported in  Refs.~\onlinecite{raschke_2004} and \onlinecite{Huang_2003}. The first peaks do appear slightly too low and to narrow compared to the simulations. Furthermore, it can be seen also that for this type of molecules, the effect of the solute charges on the solvation structure is relatively mild. From now on, we will focus on the hydrophobic solvation per se and consider only purely hydrophobic solutes, with no charge distribution. In that case the functional depends on $n(\rr)$ only (incidentally, convergence is even much faster since the angular dependence may be omitted).   

\begin{figure}
    \includegraphics[width=8.5cm]{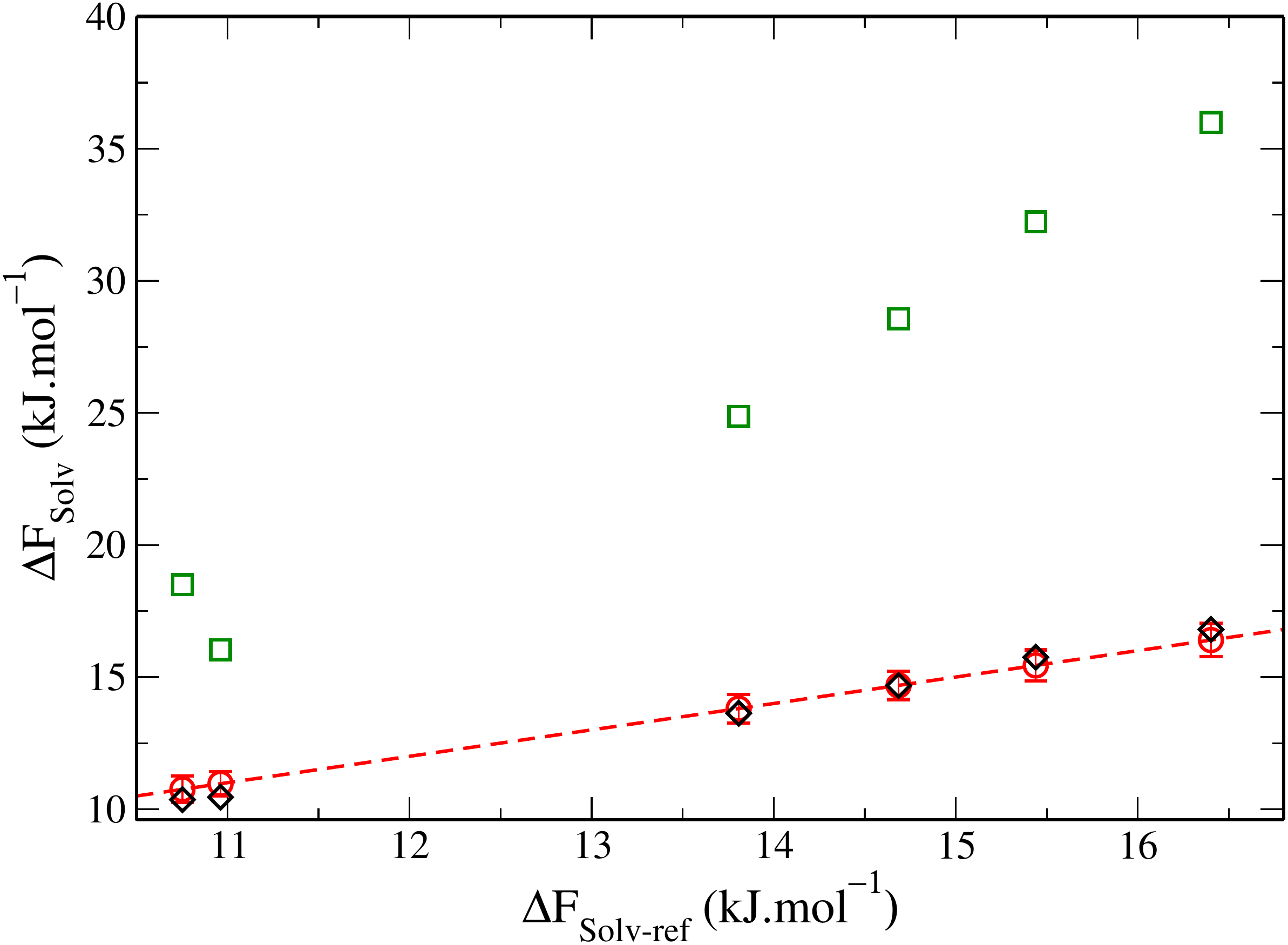}
    %~ \includegraphics{FIGURES/fn.pdf}
    \caption{
        \label{fig:free_energies_alcanes}
        Comparison of solvation free-energies values for a set of alcanes:\\
        The figure correlates the solvation free-energy values for the 6 first straight-chain alkanes (from methane on the left to hexane on the right).
        MDFT results without corrections are shown in green squares and MDFT with hard-sphere bridge and Van-der-Waals correction results are shown in black. The reference MD results, computed by Ashbaugh {\em et al.}\cite{ashbaugh_hydration_1998} are shown in red, with the reported error bars. Note that the results of MDFT with HSB but without Van-der-Waals correction are also very satisfying\cite{levesque12_1}. 
               }
\end{figure}

In that category, Figure~\ref{fig:free_energies_alcanes} compares the solvation free energies obtained for a series of  neutral alcane-like  molecules (methane to hexane) to the MD results reported for the same model by  Ashbaugh {\em et al.} using thermodynamic integration techniques\cite{ashbaugh_hydration_1998}. The first lesson is that without addition of any bridge term in eq.~\ref{eq:Fn_exc}, the MDFT solvation free-energies appear too large by a factor $\sim 2$. The addition of the hard-sphere bridge functional of eq.~\ref{eq:Fn_bridge} with a radius of $1.27$~\AA  ~is able to bring those values within the MD  error bars for the whole series. On the other hand, the correction by the macroscopic hydrophobic solvation term of eq.~\ref{eq:Fn_exc_2} has very little effect for solutes of that size.

\begin{figure}
    \includegraphics[width=8.5cm]{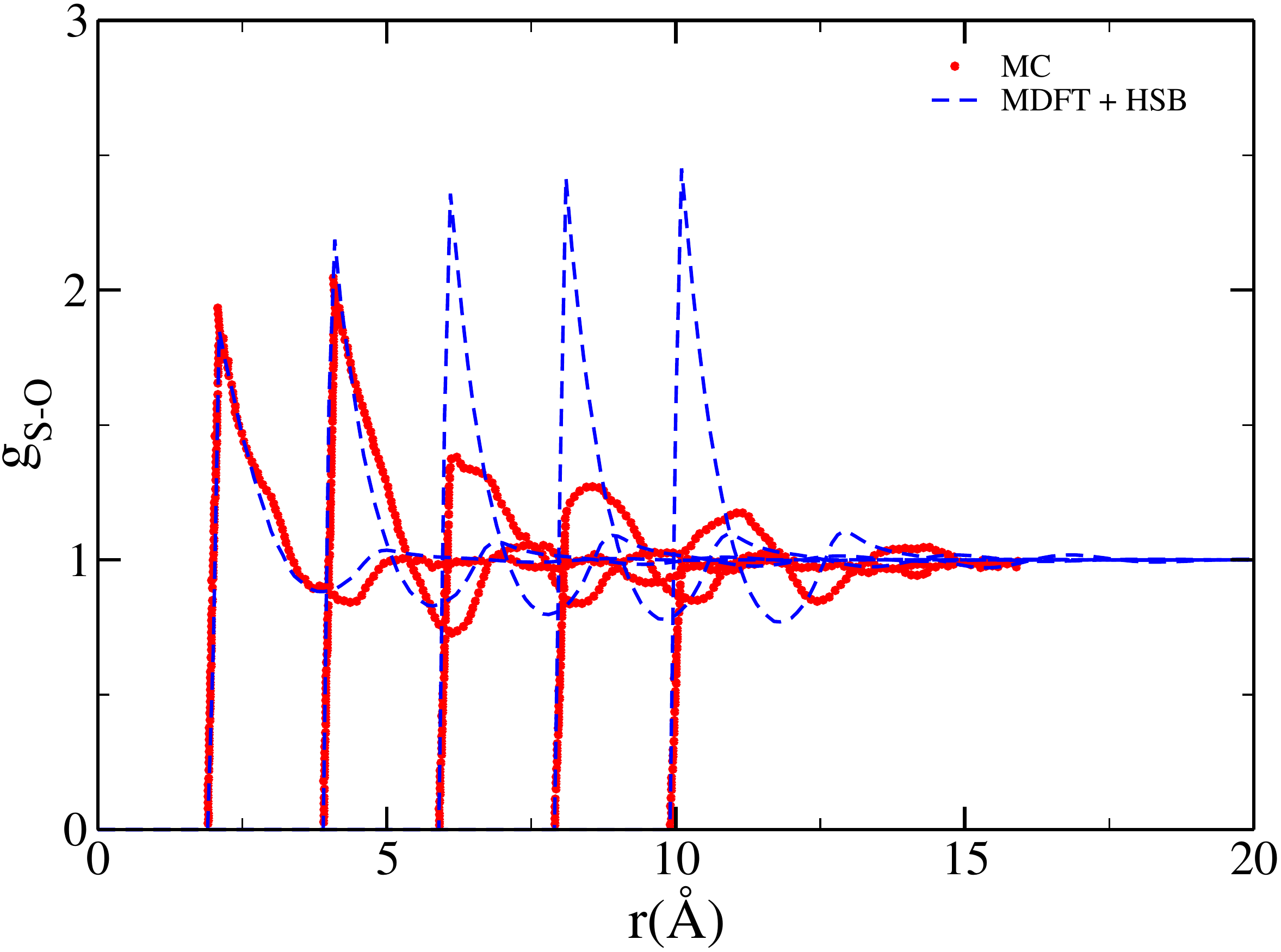}\\
    \vspace{2cm}
    \includegraphics[width=8.5cm]{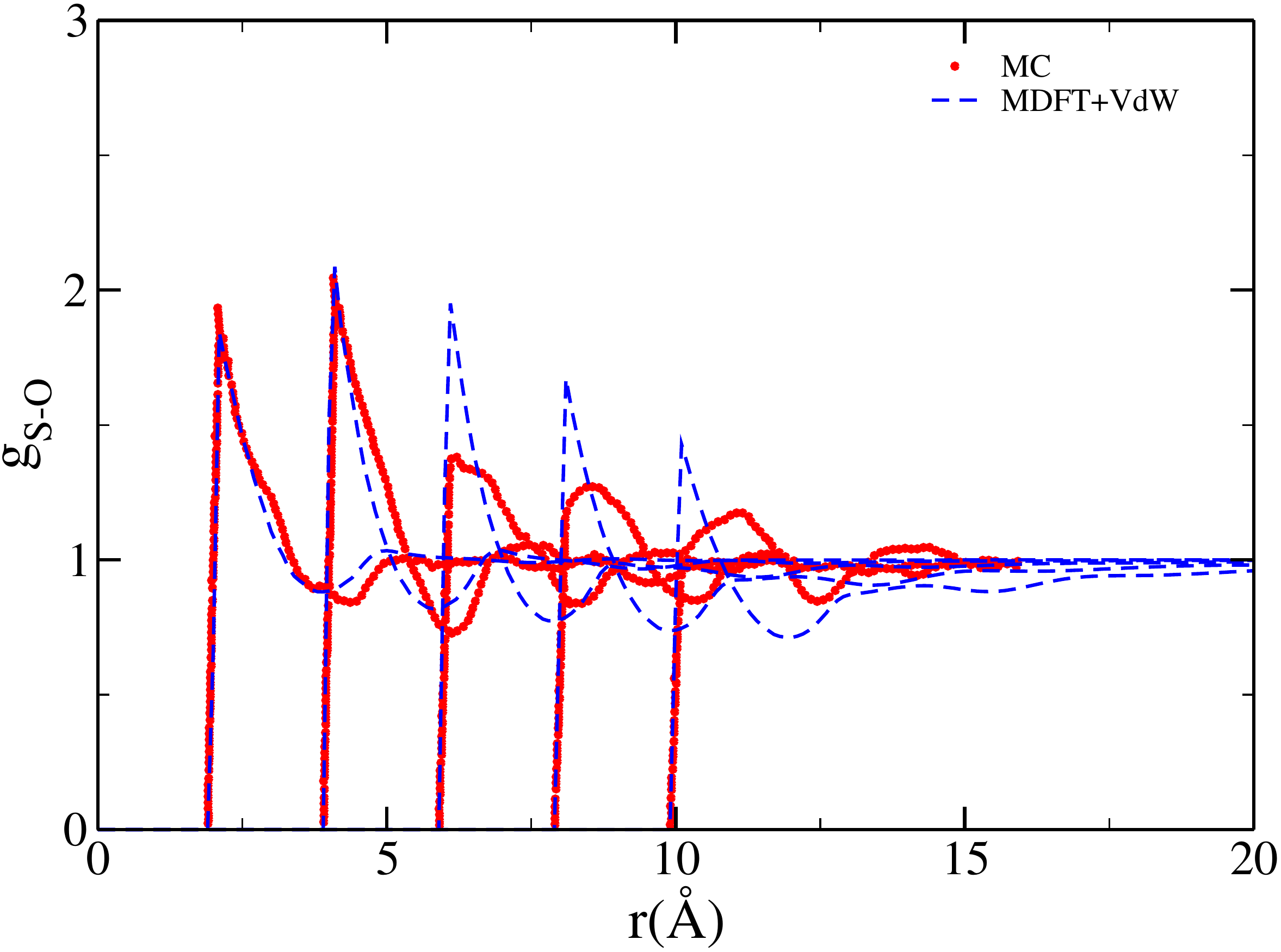}
    \caption{
        \label{fig:gr_Chandler}
        Solute-water radial distribution function for a growing hard sphere of radius $R$:\\Results obtained by MDFT minimization using only a hard-sphere bridge term (eqs~\ref{eq:Fn_exc}-\ref{eq:Fn_bridge}) (blue dashed line, top figure) or hard-sphere bridge term plus long range Van-der-Waals correction (eq.~\ref{eq:Fn_exc_2}) (blue dashed line line, bottom figure). The Monte-Carlo results of Huang and Chandler\cite{huang02} are shown in red lines.       
        }
\end{figure}

\begin{figure}
    \includegraphics[width=8.5cm]{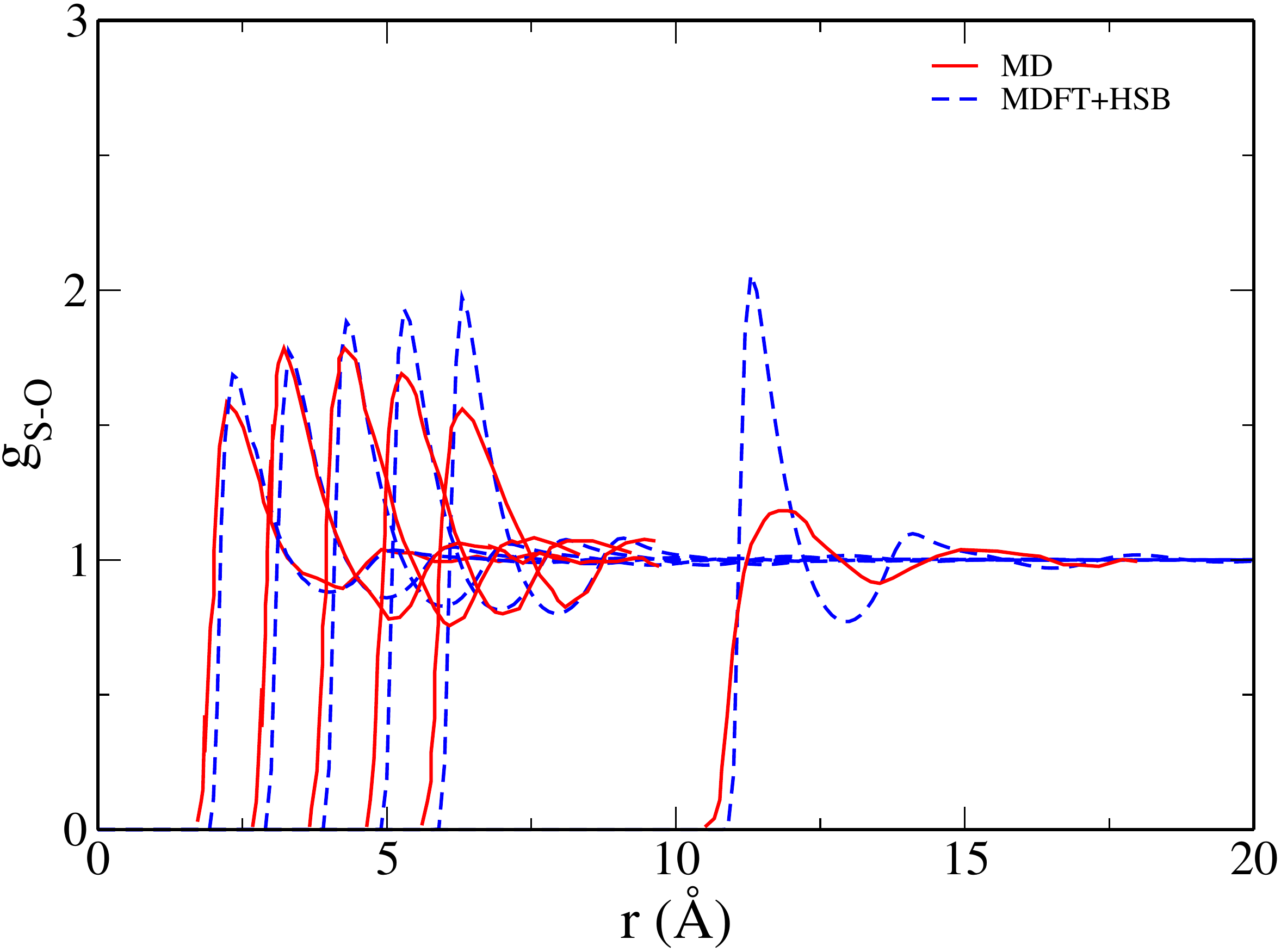}\\
    \vspace{2cm}
    \includegraphics[width=8.5cm]{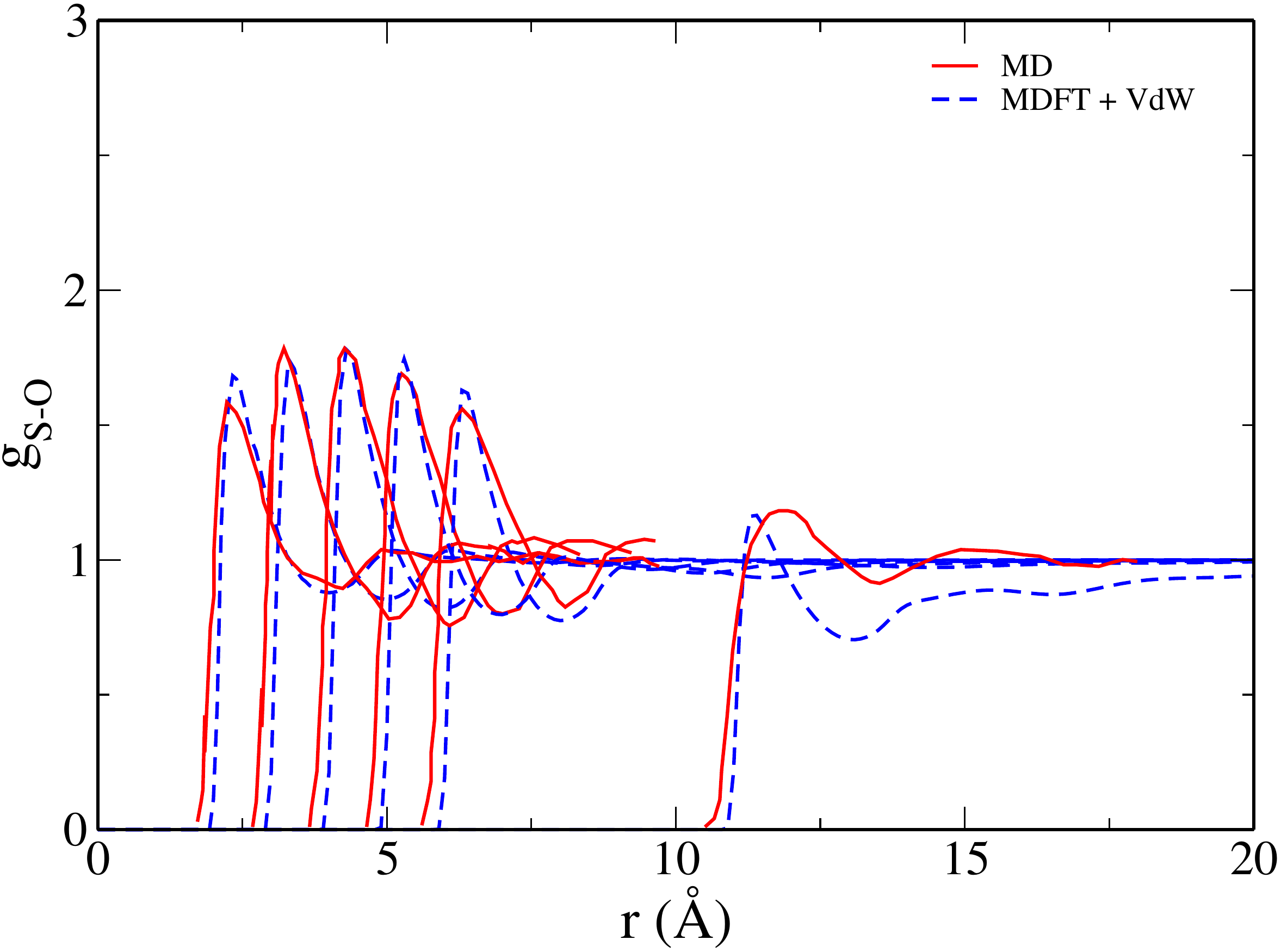}
    \caption{
        \label{fig:gr_Hansen}
        Same as Figure~\ref{fig:gr_Chandler} for a growing soft-repulsive sphere (see text) of radius $R$ and comparison to the MD results of Dzubiella and Hansen\cite{dzubiella04} in red.         }
\end{figure}

Following Lum, Chandler and Weeks\cite{lum99} and the following work of Chandler and collaborators\cite{tenwolde01,huang02,varilly11},  we next consider the hydration of a hard sphere with an increasing radius going from  \angstrom s to   a couple of nanometer.  The corresponding water densities computed by MDFT are plotted in Fig.~\ref{fig:gr_Chandler} and compared to the Monte-Carlo results of Huang and Chandler\cite{huang02}. It is seen that, when keeping the hard-sphere bridge functional, but omitting the long-range hydrophobicity correction, the densities are correct for the smallest radii up to $R=4$~\AA~but do not account for the maximum of density at contact for that value nor the further  decrease and broadening of the peak with increasing HS  radius that are   observed  in the simulations. As in Gaussian field approaches, this striking  non-monotonic behavior is recovered by incorporation of the long-range hydrophobicity term. The theory is still far from perfect:  the form of the peaks for $R>4$~\AA ~is not quite reproduced and they appear somewhat too high and too narrow with respect to the MC data. It seems also that depletion is overestimated at intermediate distances between 10 and 20~\AA. To supplement that study, we borrowed the results of the MD simulations of Dzubiella and Hansen\cite{dzubiella04} and performed the same type of calculations  for a soft-repulsive sphere creating an external potential   $\Phi(\rr) = k_BT/(R -1)^{12}$. 
The conclusions drawn from Fig.~\ref{fig:gr_Hansen} are identical, with  an even better (and almost perfect) reproduction of the solvent structure up to $R=6~$\AA ~but the same overestimated depletion for the largest sphere. 

\begin{figure}
    \includegraphics[width=8.5cm]{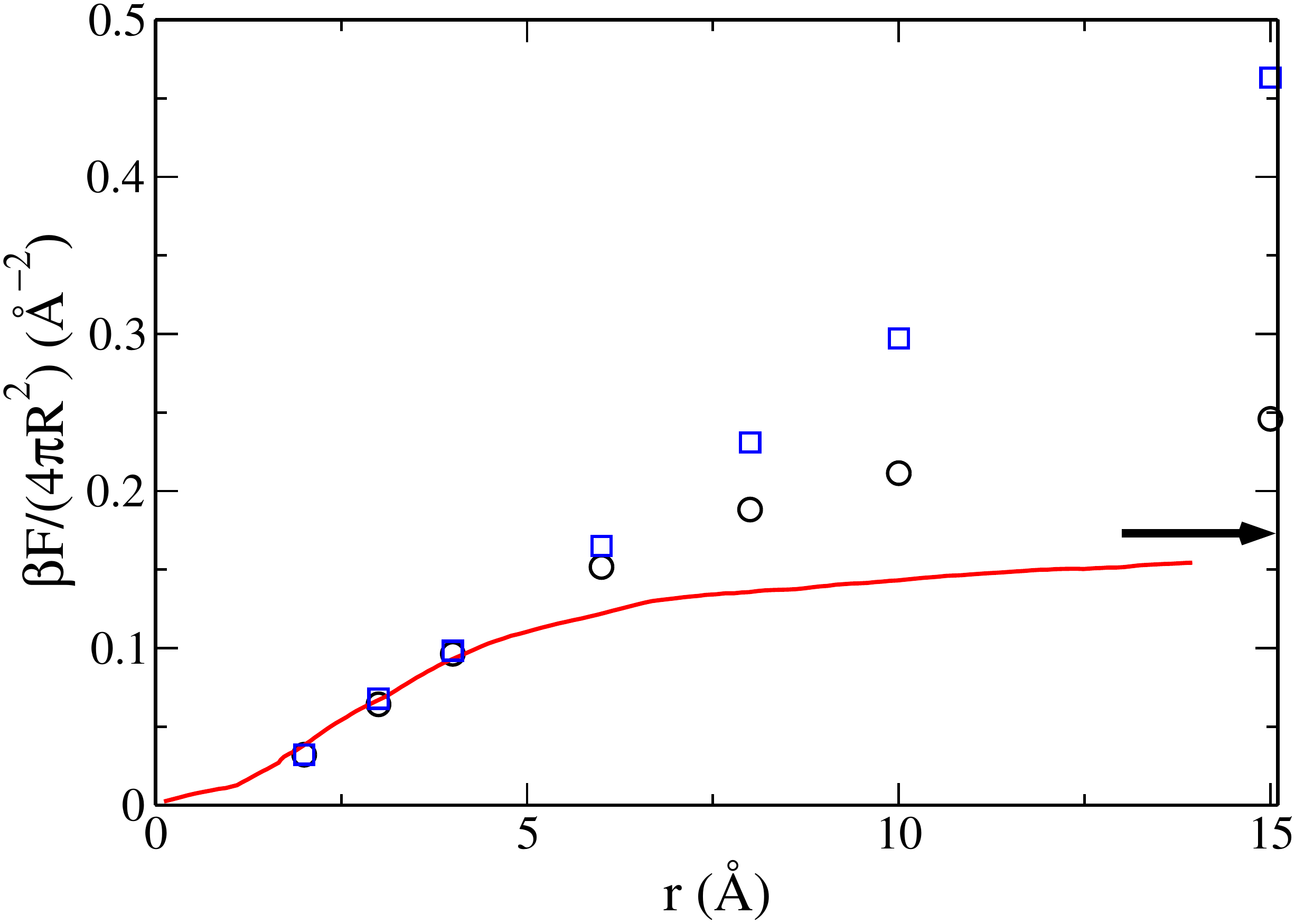}
    \caption{
        \label{fig:free_energy_Hansen}
        Solvation free energy of soft-repulsive spheres (see text) on different radius $R$:\\
        The results are obtained by MDFT minimization using only the hard-sphere bridge term of eqs~\ref{eq:Fn_exc}-\ref{eq:Fn_bridge} (blue squares) or the hard-sphere bridge term plus long range Van-der-Waals correction of eq.~\ref{eq:Fn_exc_2} (black circles). The reference MD results of Dzubiella and Hansen\cite{dzubiella04} are shown in red. The arrow
        indicates the expected limit which is the experimental value of the liquid-vapor surface tension of water at ambient conditions\cite{Lemmon}.        }
\end{figure}

To exploit further Dzubiella-Hansen's study, we also report in Fig.~\ref{fig:free_energy_Hansen} the solvation free-energies per solute surface unit as a function of the hydrophobic sphere radius, as  computed by MDFT minimization, and  we compare them to the MD results, obtained by thermodynamics integration. Such plot emphasizes the fact that $\beta \Delta F_{solv}$ should be proportional to the solute volume for small sub-nanometer radii, and become proportional to its surface for macroscopic solutes, with a proportionality coefficient equal to the liquid-gas surface tension, $\gamma$. This is indeed what the MD results show. Our MDFT approach with hard-sphere bridge functional and long-range hydrophobic correction is quite accurate at small radii, as already shown earlier for small hydrophobic solutes, and it does give the correct physical behavior for larger solutes, with a qualitative change between volume to surface behavior occurring between 5 and 10~\AA. Nevertheless,  with the current parametrization, we get a plateau value that is too high. The values adopted here for the key parameters, namely  $R_0$ (hard sphere reference fluid), $a$ (Van-der-Waals energy parameter) and $m$ (interfacial energy term), represent to our eyes the best compromise between reproduction of structure in Figs~\ref{fig:gr_Chandler} and \ref{fig:gr_Hansen} and correct reproduction of solvation free-energy in Fig.~\ref{fig:free_energy_Hansen}. Within the constraints that we imposed to our theory, in particular the Van-der-Waals approximation in eq.~\ref{eq:F_exc_expansion} that limits the expansion of the long-range attractive excess free-energy to the leading orders, improving the long-range behavior in Fig.\ref{fig:free_energy_Hansen} implies to over-emphasize the depletion phenomenon observed for the largest radius in Figs~\ref{fig:gr_Chandler} and \ref{fig:gr_Hansen}. The interfacial free energy brought by the square gradient term in eqn.~\ref{eq:F_exc_expansion}  is indeed physically very important and expected  at large scales.  The present parametrization does not give yet proper justice to it,  with an interfacial profile  that seems too large and an energetic contribution that is too low.

We see two ways out of this limitation. First the expansion in eq.~\ref{eq:F_exc_expansion} could be extended at higher order, in order to allow for more flexibility in the  definition of the double-well macroscopic functional $F_{VdW}(\bar{n})$ of Fig.~\ref{fig:fn} and 
eqs~\ref{eq:F_n_int}-\ref{eq:F_VdW}, and to make it possible to play independently  with the location of the first minimum  and the barrier height. Secondly, we think there are still some clues to be clarified on the link between the MD  structures and free-energies, which were computed in the canonical ensemble, and the MDFT calculations that are performed by essence in the grand-canonical ensemble.

\section{Conclusion}

In this paper, we have presented an extension of the molecular density functional theory of water that was introduced in Refs~\cite{zhao11,jeanmairet13} that is able to describe hydrophobic solvation from sub-nanometer to nanometer and, supposedly, up to macroscopic length scales.  This required to modify the density part in the water functional proposed in Ref.~\cite{jeanmairet13} by including a hard-sphere bridge functional that acts at all scales and then tuning the second-order attractive term at long length scales in order to match a  consistent Van-der-Waals theory of liquid-vapor coexistence at those scales. Such procedure is largely inspired  by the LCW theory of hydrophobicity and it takes care of the possibility of a dewetting transition caused by the presence hydrophobic solute at those scales.  
The specificity of our MDFT approach with respect to that of Lum, Chandler, and Weeks should be stressed however. In the straight LCW theory, a macroscopic equation such as eq.~\ref{eq:F_exc_expansion} is written for a coarse grained density and a linear response approximation is adopted for the microscopic density around the coarse-grained one. The mean microscopic density and coarse-grained density are coupled through a linear mean force term. Our formulation starts from a general free-energy expression for the microscopic density $\rhon$, and the exact decoupling of this density into a short range, $\rhon - \rhonbar$, and long range part, $\rhonbar$, defined as a convolution of the microscopic density $\rhon$ by a coarse-graining function $G(r)$ (a  weighted density  approach in the DFT language\cite{sun01}; see eq.~\ref{eq:rhobar}-\ref{eq:14}). The excess free energy  can then be decomposed also into a short-range  and long-range, Van der Waals part, the later  involving two parameters that can be tuned phenomenologically. As  for the  ideal  and the external free energies, they  remain defined at the fully microscopic level. Overall, $\rhonbar$ being a function of $\rhon$, the complete functional is  minimized with respect to the microscopic density itself, so that the theory is fully self-consistent. As a technical drawback, the fineness of the grid to be employed is imposed by the microscopic density, not the coarse-grained one,  and should remain at the sub-Angstrom level.

\begin{figure}
    \includegraphics[width=8.5cm]{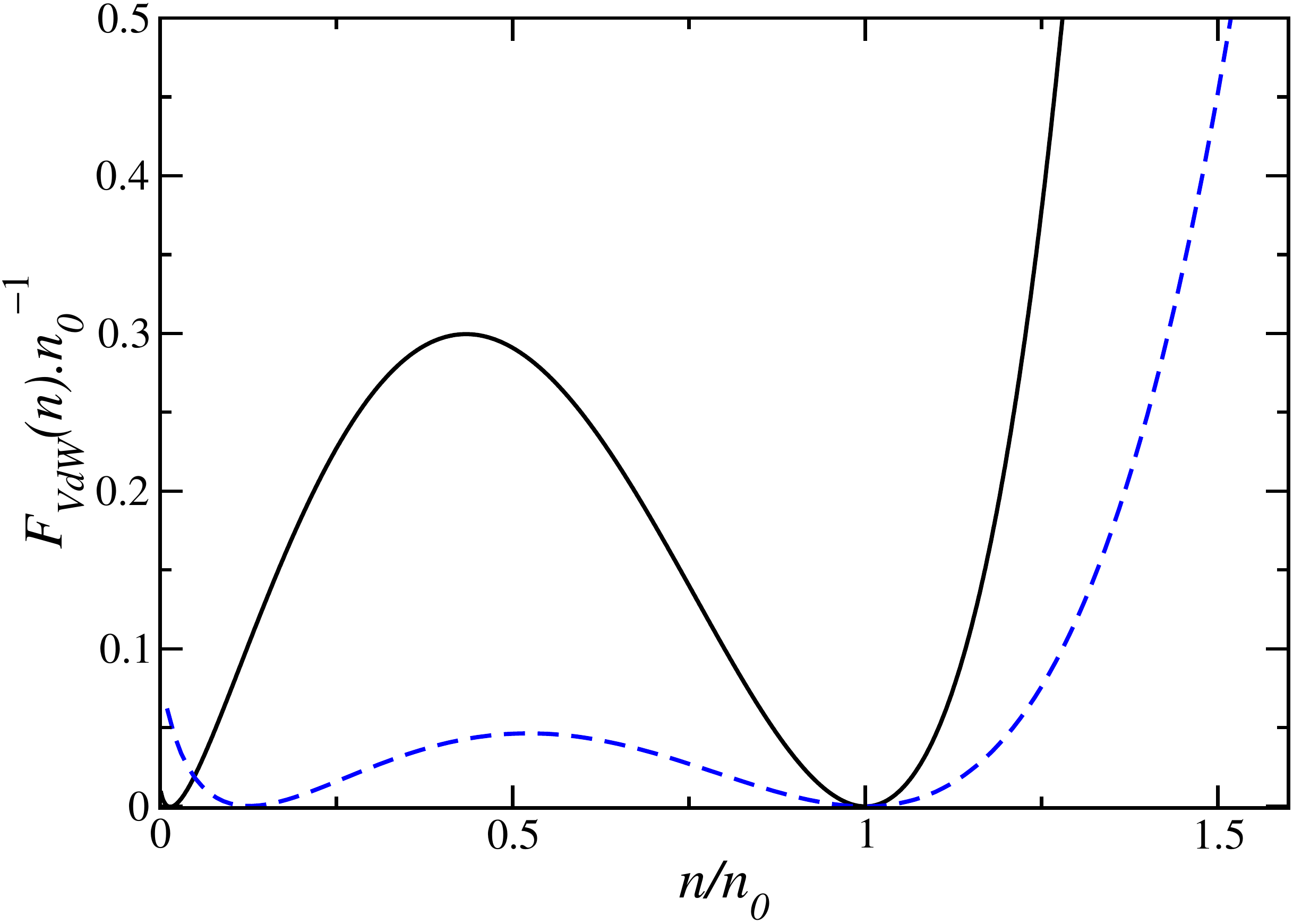}
    \caption{Van-der-Waals functions for the coarse-grained densities:\\
    Those functions of eq.~\ref{eq:F_VdW} describe  the (near) coexistence of liquid and vapor for the coarse-grained densities. The blue dashed line and black line correspond to two different radius for the hard-spheres of the reference fluid:  $R_0 = 1.25$~\AA~ and $R_0 = 1.4$~\AA~, respectively.
        \label{fig:fn}
        }
\end{figure}

The MDFT functional minimization without the above correction was shown to yield the correct water structure around microscopic hydrophobic solutes, including or not the charge distribution that describes those molecules in force fields. The corrected functional contains the appropriate physics to describe also the non-monotonic behavior of the first solvation peak that is observed in molecular simulations when increasing the hydrophobic solute size, as well as the transition from a volume-driven to surface-driven behavior for the solvation free energy. The theory is not perfect at this stage and, in particular,  the value observed for the plateau of curve giving the free energy per exposed surface area  as function of solute size is too high, and  it does not point to the correct value of the water liquid-vapor surface tension. This could be improved by making the model  a little bit more complex and introducing terms of order $\Delta \rhonbar^3$ and  $\Delta \rhonbar^4$ in the Van-der-Waals expansion of eq.~\ref{eq:F_exc_expansion} in order to be able to modulate independently the location of the gas-phase minimum and the free-energy barrier height in the function $F_{VdW}(\bar{n})$ of eq.~\ref{eq:F_VdW} and Fig.~\ref{fig:fn}. This should be combined with the fundamental interfacial $\nabla n(\rr)^2$ term to yield a correct description of the water liquid-gas interface, in terms of  both width and  energy.  
 We mention also that, as in Refs~\cite{dzubiella04,roundy13}, the problem of hydrophobic spheres interactions should be studied too.
 
Concerning numerical efficiency, the incorporation of both the hard-sphere bridge functional or the coarse-grained Van-der-Waals corrections, although increasing the number of FFT's to be performed,  do not cause a huge increase of the calculations with respect to the straight HRF approximation, as it was implemented to treat extended molecular systems, such as clay-water interfaces\cite{levesque12_2}. The cost of the calculation scales basically as $N^3$, the number of grid points\cite{levesque12_1}; with a recommended grid spacing of 3-4 points per angstrom, systems up to 80 $\AA^3$ can  easily be contemplated. The application of the present multi-scale MDFT  approach to the water structure around biomolecules and the elucidation of hydrophobic versus hydrophilic solvation is presently under way.

Finally, in the present functional approach, the coupling between the density and polarization fields occurs only through the ideal term in eq.~\ref{eq:FP}, so that a solute with zero charge implies a vanishing polarization. We have not yet introduced  a  cross term in the excess free-energy that should describe the direct coupling between the density field $n(\rr)$ and the  charge density field $\nabla \cdot \PP(\rr)$. We estimated such coupling to be of second order due to the smallness of the corresponding terms in the rotational invariants expansion of the angular-dependent direct correlation function of bulk water (the symmetry of such coupling is expressed in the $c_{101}(r)$ function, which we found very small with respect to other components such as $c_{110}(r)$ or $c_{112}(r)$; see Ref.~\onlinecite{zhao13}).  Nevertheless the question of this coupling remains important and its implications on hydrophobic solvation properties are worth being examined more  quantitatively in a close continuation of this work.

%~ \begin{acknowledgments}
%~ XXX
%~ \end{acknowledgments}

\newpage

\bibliographystyle{apsrev4-1}
\bibliography{main}

\newpage

\clearpage
%fig2

%fig3

%fig4

%fig5

%fig6

%fig7

%fig8

\end{document}